\documentclass[preprint,10pt]{elsarticle}

\usepackage{graphicx}
\usepackage{amsmath}
\usepackage{bm}

\biboptions{square,sort&compress}
\journal{High Energy Density Physics}

\begin{document}
\begin{frontmatter}
\title{Opacity of Shock-Heated Boron Plasmas}
\author{W. R. Johnson}
\ead{johnson@nd.edu}
\address{University of Notre Dame, Notre Dame, IN 46556}
\author{J. Nilsen}
\ead{nilsen1@llnl.gov}
\address{Lawrence Livermore National Laboratory, Livermore, CA 94551}
\begin{abstract}
Standard measures of opacity, the imaginary part of the atomic scattering factor $f_2$ and the x-ray  mass attenuation coefficient $\mu/\rho$,
are evaluated in shock-heated boron, boron carbide and boron nitride plasmas.
The Hugoniot equation, relating the temperature $T$ behind a shock wave to the compression ratio $\rho/\rho_0$ across the shock front, is used in connection with the plasma equation of state to determine the pressure $p$, effective plasma charge $Z^\ast$ and the K-shell occupation in terms of $\rho/\rho_0$. Solutions of the Hugoniot equation (determined within the framework of  the generalized Thomas-Fermi theory) reveal that the K-shell occupation in low-Z ions  decreases rapidly from 2 to 0.1 as the temperature increases from 20eV to 500eV; a temperature range in which the shock compression ratio is near 4.
The average-atom model (a quantum mechanical version of the generalized Thomas-Fermi theory) is used to determine K-shell and continuum wave functions and the photoionization cross section for x-rays in the energy range $\omega=1$~eV to 10~keV, where the opacity is dominated by the atomic photoionization process.
For an uncompressed boron plasma at $T=10$~eV, where the K-shell is filled, the average-atom cross section,  the atomic scattering factor and the mass attenuation coefficient are all shown to agree closely with  previous  (cold matter) tabulations \cite{yeh:85,hen:93,hubb:95}.
For shock-compressed plasmas, the dependence of  $\mu/\rho$ on temperature can be approximated by scaling previously tabulated cold-matter values by the relative K-shell occupation; however, there is a relatively small residual dependence arising from the photoionization cross section.  Attenuation coefficients $\mu$ for a 9 keV x-ray are given as functions of $T$ along the Hugoniot for B, C,  B$_4$C and BN plasmas.
\end{abstract}

\begin{keyword}
52.50.Lp: Plasma production and heating by shock waves and compression
\sep
56.65.Rr: Particle in cell method
\sep
52.70.-m: Plasma diagnostic techniques
\sep
52.25.Os:	Emission, absorption, and scattering of electromagnetic radiation.
\end{keyword}
\end{frontmatter}

\section{Introduction}
The aim of this paper is to examine the opacity of the hot dense plasma behind an intense shock wave impinging on matter at standard temperature and pressure. For simplicity,  the present study is restricted to boron, carbon and two boron compounds, boron carbide and boron nitride; however, the methods developed herein are applicable to other light-ion plasmas such as polystyrene. 

A brief review of shock waves and the Hugoniot equation, which governs the thermodynamic properties of the plasma behind a shock front, is given first. This discussion is 
followed by a review of the generalized Thomas-Fermi (TF) theory of Feynman, Metropolis and Teller \cite{FMT:49}, which is used to determine the energy per ion and pressure on the two sides of shock front in terms of the respective temperatures and densities. With such information, the Hugoniot equation can be solved to give the temperature $T$, pressure $p$, effective ionic charge $Z^\ast$ and the occupation of the ionic K-shell as functions of the compression ratio $\rho/\rho_0$. 

The K-shell occupation, which is a critical factor in determining the plasma opacity, is determined by solving the Schr\"{o}dinger equation in the TF potential.  Such a procedure is
based on the assumption that the semi-classical TF potential is a good approximation to a proper quantum-mechanical potential. To test this assumption, the K-shell occupation is
re-evaluated using a quantum-mechanical average-atom cell model to describe the plasma. Differences of 10\% or less are found between the semi-classical and quantum-mechanical determinations of the K-shell occupation numbers.

Following this comparison, we introduce the (complex) atomic scattering factor $f(\theta)$ together with the relation between the index of refraction $n$ in a medium and the forward scattering factor $f(0)$. The forward-scattering factor $f(0)$ is tabulated as a function of photon energy by Henke, {\it et al.} \cite{hen:93} for (cold) elements $1\leq Z \leq 92$. The imaginary part $f_2$  ($f(0) = f_1- i f_2$),
 which is a function of the photon energy $\omega$, is responsible for the exponential decay of the intensity of an electromagnetic wave. In the x-ray region beyond the K-shell threshold $f_2$ is proportional to the atomic bound-free photoionization cross section $\sigma_\text{bf}(\omega)$.

The average-atom model is used to evaluate the photoionization cross section $\sigma_\text{bf}(\omega)$ for energies above the K-shell threshold ($\approx 200$ eV for boron) to 10 keV.
It should be noted that average-atom models in various guises have been used previously to study photoionization and opacity in
warm- and hot-dense plasmas \cite{Yu:02, BR:08, PB:11, MW:13, HC:15, JH:16, SF:17, HC:17, SK:18, TN:18, PB:18}. 
The photoionization cross section calculated using the present model for a 
boron plasma at temperature $T=10$~eV and density $\rho=2.463$~g/cc, where the K-shell is
filled, is in close agreement with the boron K-shell cross section given in the tabulation of Yeh and Lindau \cite{yeh:85}. With the aid of the relation between
$f_2$ and $\sigma_\text{bf}$, one determines $f_2$ for shock-compressed plasmas\text.
Close agreement is found between the resulting average-atom calculations of $f_2$
and values in the Henke tabulation for a boron plasma at temperature $T=10$~eV and density $\rho=2.463$~g/cc.

Finally, the x-ray mass attenuation coefficient $\mu$, tabulated for cold matter by Hubbell and Seltzer \cite{hubb:95}, is introduced and expressed in terms of the $\sigma_\text{bf}$.
The coefficient $\mu$, which is the reciprocal of the $1/e$ x-ray intensity falloff distance, 
is proportional to the plasma density; the ratio $\mu/\rho$ for 9~keV x-rays is found to be strongly dependent on the K-shell occupation in the 20-500~eV temperature range and weakly
dependent on the remaining factors governing the photoionization cross section.  
Plots are presented illustrating $\mu(\omega)$ for a 9~keV He-like Zn x-ray as a function of $T$ behind a shock moving into cold B, C, B$_4$C and BN plasmas.

\section{Shock Waves and the Hugoniot Equation}
\begin{figure}
\centerline{\includegraphics[scale=0.35]{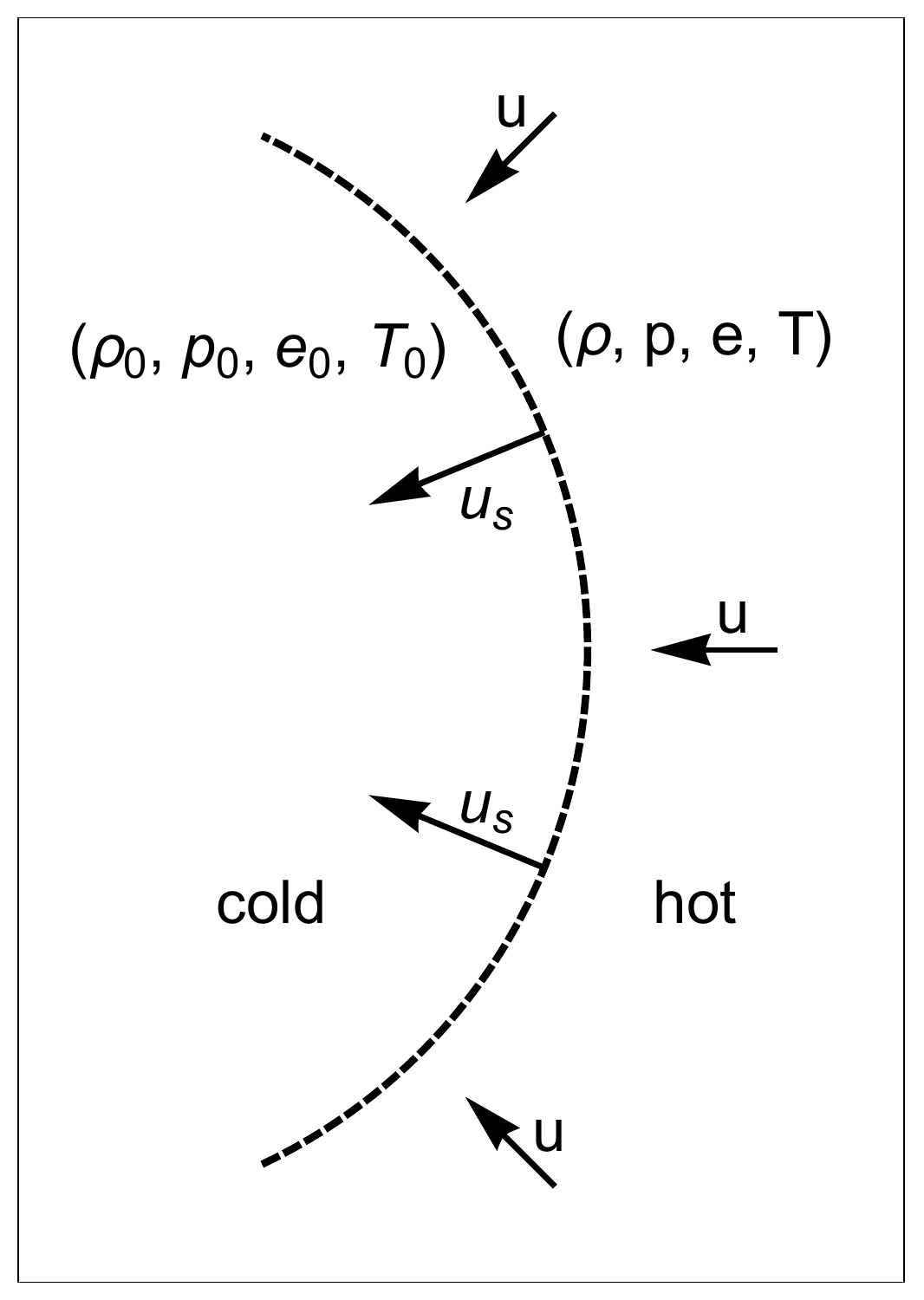}}
\caption{\label{fig1}  A shock wave, illustrated by the dashed arc, moves with velocity $u_s$ into a cold medium,
leaving behind a hot compressed medium moving with velocity $u$.}
\end{figure}

A shock wave, illustrated in Fig.~\ref{fig1}, moving with velocity $u_s$ into a (cold) medium at rest having pressure $p_0$, density $\rho_0$, energy per ion $e_0$ and temperature $T_0$  leaves behind a hot compressed medium with pressure $p$,
density $\rho$, energy $e$ and temperature $T$ moving with velocity $u$. These quantities are subject to the conservation laws (formulated for a plane shock wave in a reference frame moving with the shock):
\begin{align*}
\rho (u_s -u) &=\    \rho_0  u_s                                             & \text{Continuity} \\
p  + \rho (u_s-u)^2   &=\   p_0 + \rho  u_s^2                         & \text{Conservation of Momentum} \\
e + \frac{1}{2} (u_s-u)^2  &=\    e_0  + \frac{1}{2}  u_s^2        &\text{Conservation of  Energy}
\end{align*}
Determining $u_s$ and $u$ from the first two equations leads to the following reformulation of the law of conservation of energy:
\begin{equation}
H(T,\rho) =  (e-e_0) - \frac{1}{2} (p+p_0) \left( \frac{1}{\rho_0}-\frac{1}{\rho} \right) = 0 . \label{heqn}
\end{equation}
Eq.(\ref{heqn}), referred to as the Hugoniot equation, can be solved to obtain both the temperature $T$ and the pressure $p$ in the shocked region in terms of the density ratio $\rho/\rho_0$,
assuming that the thermodynamic properties of the un-shocked medium are known.

 The density $\rho$ and pressure $p$ in the shocked medium can be determined experimentally as illustrated in Ref.~\cite{SK:18} from knowledge of the density
$\rho_0$ and  pressure $p_0$ in the un-compressed medium and measurements of $u$ and $u_s$ using the following relations that can be inferred from the conservation laws:
\begin{align}
\rho/\rho_0 &=\, u_s/(u_s-u) \\
          p      &=\,  p_0+ (1-\rho_0/\rho) \rho_0 u_s^2 .
\end{align}

In the present work, we determine the contribution of plasma electrons to $H(T,\rho)$ using the generalized Thomas-Fermi theory of  Feynman, Metropolis and Teller \cite{FMT:49} and determine contributions of ions to  $H(T,\rho)$ using the ideal gas law.

\section{Generalized Thomas-Fermi Theory}

The generalized Thomas-Fermi theory  is a well-known method for evaluating the electron pressure $p$, chemical potential $\mu$, free electron number density $n_e$ and energy $e$ in terms of temperature $T$ and density $\rho$. In this theory, a neutral plasma consisting of ions of nuclear charge $Z$ and electrons is divided into neutral Wigner-Seitz (WS) cells of volume $V_\text{WS}=A/(N_A \rho)$, where $A$ is the atomic weight of the ion and $N_A$ is Avogadro's number. Inside the WS cell, electrons are assumed to move in a self-consistent potential:
\begin{equation}
 V(r) = -\frac{Z}{r} + \int_{r'\leq R} \! \!  \frac{n_e(\bm{r}')d^3r'}{|\bm{r}-\bm{r}'|},   \label{e4}
\end{equation}
where  $R$ is the radius of the WS sphere. The electron density $n_e(r)$ in Eq.~(\ref{e4}) is given by the semi-classical expression:
\begin{equation}
 n_e(r) = \frac{1}{\pi^2} \int_0^\infty \frac{p^2 dp }{1 + \exp[ (p^2/2m + V(r) - \mu)/kT ]},
 \label{netf}
\end{equation}
where $p$ represents electron momentum.
The chemical potential $\mu$
is determined by the requirement that the cell be neutral:
\begin{equation}
\int_{r \leq R}\! \!  n_e(r) d^3 r  = Z . \label{e6}
\end{equation}
Equations (\ref{e4}-\ref{e6}) are solved self-consistently to give the electron density inside the WS cell,
the potential $V(r)$ and the chemical potential $\mu$.  The potential $V(r)$ satisfies
\begin{equation}
  V(R) =0 \quad \text{and} \quad \frac{dV(R)}{dr} = 0 .
\end{equation}

\begin{figure}
\centerline{\includegraphics[scale=0.5]{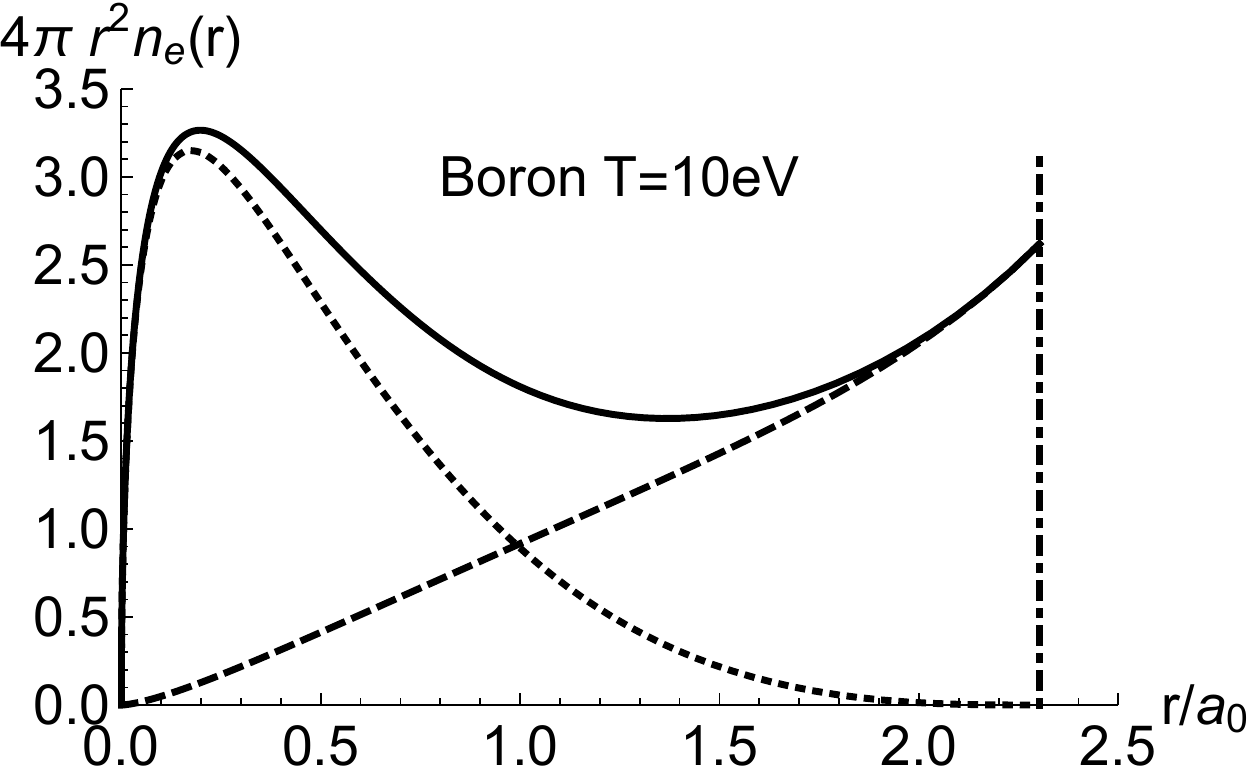}
\quad
\includegraphics[scale=0.5]{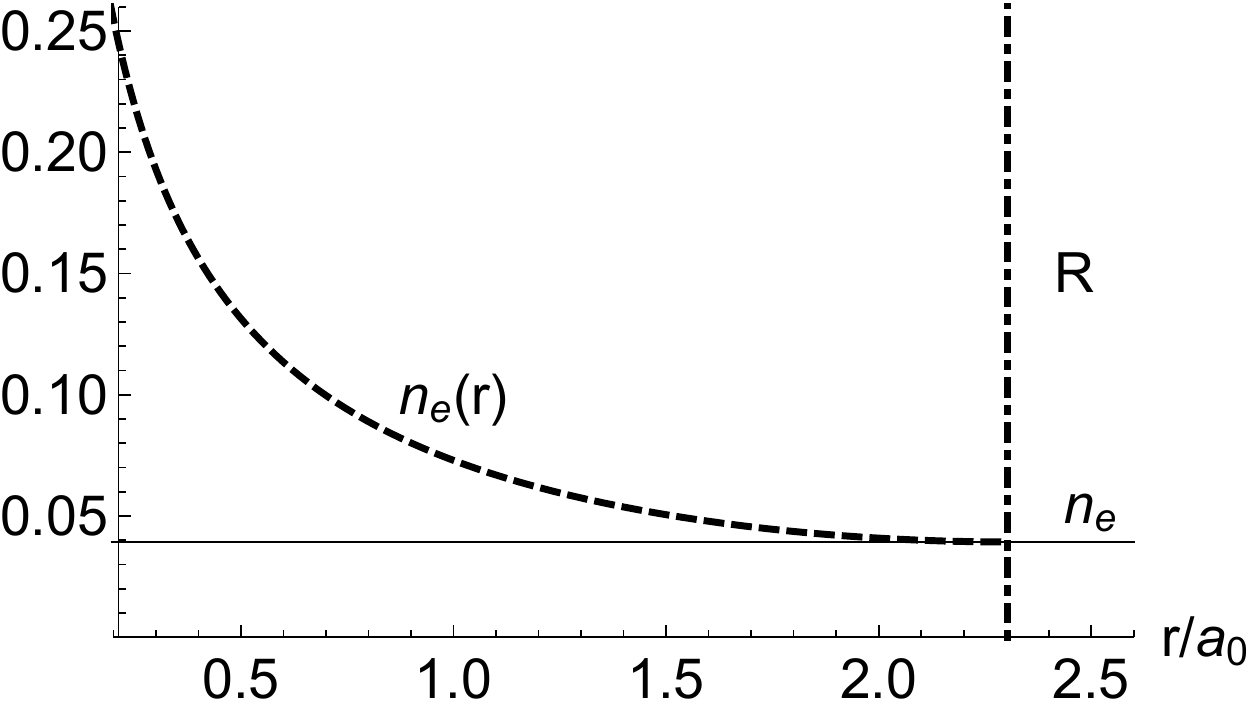}}
\caption{\label{fig2} Left panel: The solid curve shows the radial electron number density $4\pi r^2 n_e(r)$ obtained by solving the TF Eqs.(\ref{e4}-\ref{e6}) for a boron plasma at temperature
$T=10$~eV and density $\rho=2.463$~g/cc. The dotted and dashed curves represent contributions to the radial density for energies $p^2/2m+V(r)$ below and above zero, respectively. Right panel: The $p^2/2m+V(r)>0$
 (continuum) contribution to $n_e(r)$ approaches a constant value $n_e$ at the cell boundary, interpreted as the average electron density of the plasma .}
\end{figure}

The solid curve in the upper panel of Fig.~\ref{fig2} shows the radial density $4\pi r^2 n_e(r)$ for a boron plasma at $T=10$~eV and density $2.463$~g/cc. The contribution to the integral in Eq.(\ref{netf}) from energies $p^2/2m+V(r) <0$ are illustrated by the dotted curve. This contribution is the semi-classical counterpart of quantum mechanical bound states. The dashed curve shows the contribution to the radial density from energies $p^2/2m+V(r) \geq 0$. In the right panel, we divide out the factor $4\pi r^2$ from the dashed curve and show that the
continuum contribution to $n_e(r)$ smoothly approaches a constant value
$n_e = n_e(R)$ at the WS cell boundary.  The value $n_e$ is interpreted as the density of the uniform electron background in which the WS cell floats. To establish neutrality outside the WS cell, there must be a compensating ion background charge $Z^\ast = n_e/n_i$.

In the generalized Thomas-Fermi theory, the contribution of electrons to the pressure is
\begin{equation}
 p_e = \frac{(2m k T)^{5/2}}{6m\pi^2} I_{3/2}(\mu/kT) \ ,
\end{equation}
where
\begin{equation}
I_{3/2}(x) = \int_0^{\infty}\!\! \frac{y^{3/2}\, dy }
{\left[1+e^{y-x}\right]} .
\end{equation}
The electron contribution to the potential energy per ion is
\begin{multline}
 e_{\rm e-pot} =  \int_0^R\!\! 4\pi r^2 \rho(r) \left( r \frac{dV}{dr} \right) dr \\
   =  - \int_0^R\!\! 4\pi r^2 n_e(r)\ \frac{Z}{r} \ dr +  \int_0^R\!\! 4\pi r n_e(r) dr  \int_0^r\!\! 4\pi s^2 n_e(s) ds  \\
     =   e_{e-n} + e_{e-e} ,\hspace{17.5em}
\end{multline}
and the electron contribution to the kinetic energy can be obtained from the ``virial'' theorem
\begin{equation}
e_{\rm e-kin} = \frac{3}{2} p V_\text{WS} - \frac{1}{2} e_{\rm e-pot} \label{vir}.
\end{equation}

As mentioned earlier, the ideal gas laws:
\begin{align}
p_i &=\ n_i k T \\
e_i &= \frac{3}{2} k T
\end{align}
are used to determine the ion contributions to pressure and energy. This treatment leads
to some error at low temperature, but is well justified in the temperature range 20-500 eV
of primary interest in the present study.
The pressure and energy in the Hugoniot equation are the sum of the electron and ion contributions.

\begin{figure}
\centerline{\includegraphics[scale=0.65]{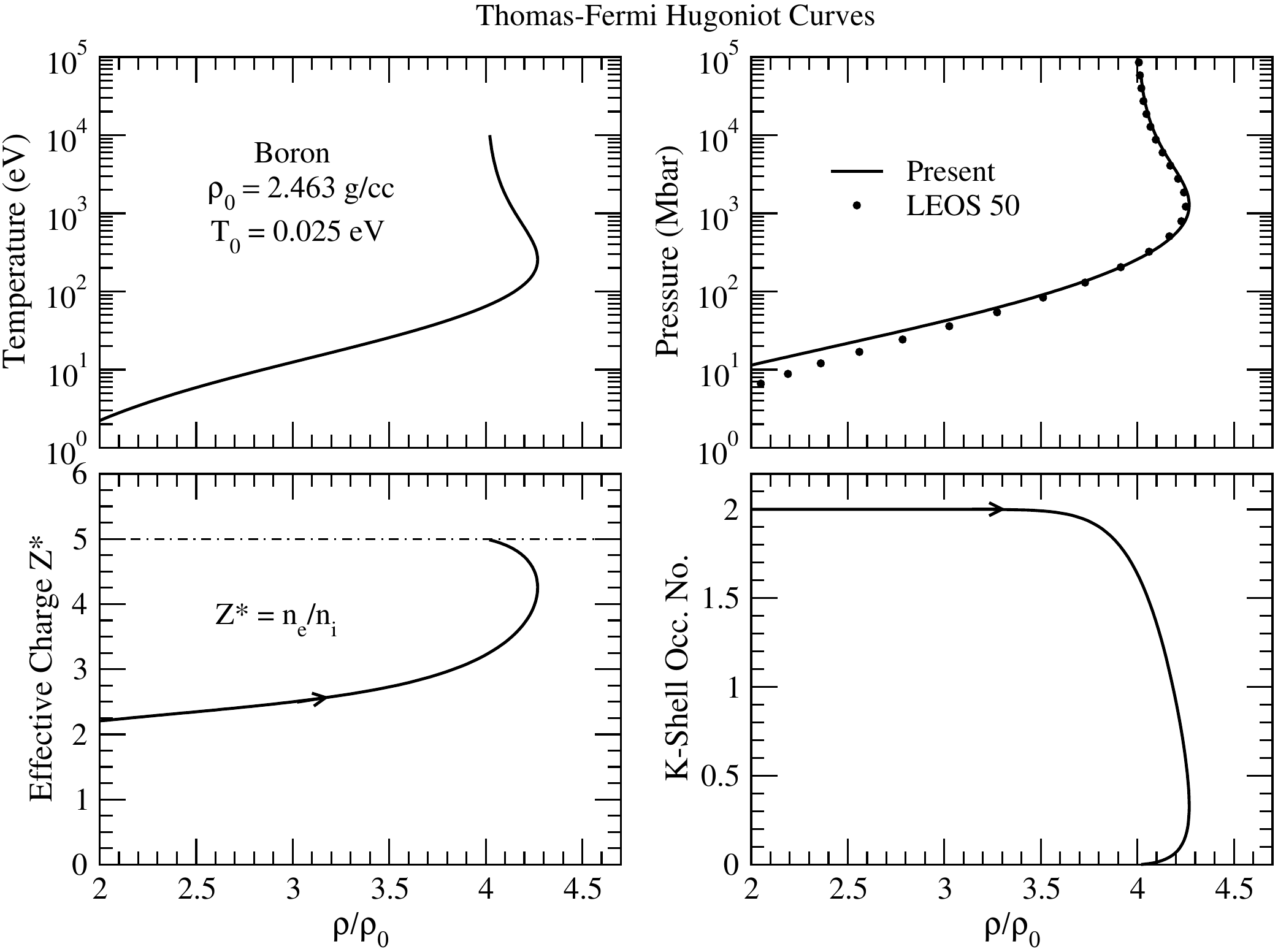}}
\caption{\label{fig3} Upper-left panel: Solution to Hugoniot equation $T(\rho/\rho_0)$. Upper right: $p(\rho/\rho_0)$ obtained with the aid of the TF equation of state is shown by the solid line and LEOS 50 data from the Livermore LEOS database is shown by dots. Lower-left: Average ionic charge of the plasma $Z^\ast(\rho/\rho_0)=n_e/n_i$.
Lower-right: K-shell occupation number determined by solving the Schr\"{o}dinger equation in the TF potential.}
\end{figure}

In the upper-left panel of Fig.~\ref{fig3}, we show the solution $T$(eV) vs.\ $\rho/\rho_0$ to the Hugoniot equation for a boron plasma initially at temperature $T_0=0.025$~eV and density $\rho_0=2.263$~g/cc.
With the aid of the plasma equation of state, the solution curve can be converted into the curve giving
$p$ as a function of $\rho/\rho_0$ shown in the upper-right panel of the figure. 
The dotted LEOS~50 curve from the Livermore LEOS database shown in the
upper right panel is obtained using an equation of state at low temperatures
in which the the TF + ideal gas expressions for pressure and energy  are modified to account for ionic bonding   \cite{MW:88,YC:98} . 
Differences between these two pressure Hugoniot curves are found to have little influence on the opacity calculations in the present work. 
On the other hand, the pressure Hugoniot curves in Fig.~\ref{fig3}  have a peak compressibility that is about 7\% 
less compressed (stiffer than) recent path-integral Monte Carlo (PIMC) and density-functional-theory molecular-dynamics (DFT-MD) calculations \cite{ZM:18} in the temperature interval where the opacity is changing rapidly!
In the lower-left panel of Fig.~\ref{fig3}, the average background ion charge $Z^\ast = n_e/n_i$
along the Hugoniot curve is shown. 
It is seen that as temperature increases $Z^\ast$ converges to 5 corresponding to complete ionization of boron ions in the plasma. 

Finally, in the lower-right panel of Fig.~\ref{fig3},  the K-Shell occupation number
inferred by solving the Schr\"{o}dinger equation in the TF potential is shown. The K-shell occupation is given in terms of the $1s$ eigenvalue $\epsilon_{1s}$ by
\begin{equation}
\text{occ no.} = \frac{2}{1+\exp[(\epsilon_{1s}-\mu)/kT]}. \label{e14}
\end{equation}

The fact that the limiting high temperature values of all four functions illustrated in Fig.~\ref{fig4} occur at $\rho/\rho_0 = 4$ is a consequence of using the ideal gas law to describe the ionic motion. 

\section{Average-Atom Model and K-shell occupation}
 A quantum-mechanical average-atom cell model is used to test the accuracy of the
 Thomas-Fermi calculation of the K-shell occupation number described in the previous paragraph.
 The average-atom cell model differs from the Thomas-Fermi theory in two essential ways. 
 Firstly, the expression for the potential $V(r)$ given in Eq.~(\ref{e4}) is modified by including
 a (local) exchange-correlation correction: 
\begin{equation}
 V(r) = -\frac{Z}{r} + \int_{r'\leq R} \! \!  \frac{n_e(\bm{r}')d^3r'}{|\bm{r}-\bm{r}'|} + V_{xc}(n_e) . \label{e44}
\end{equation}
Secondly, the semi-classical expression for $n_e(r)$ in Eq.~(\ref{netf}) is replaced by the quantum-mechanical 
expression
\begin{multline}
n_e(r) = 
 \frac{1}{4\pi r^2} \biggl[  \sum_{nl} \frac{2(2l+1)}{1+\exp[(\epsilon_{nl}-\mu)/kT]} P^2_{nl}(r)   \\
    + \sum_i \int_0^\infty \!\! d\epsilon  \frac{2(2l+1)}{1+\exp[(\epsilon-\mu)/kT]}   P^2_{\epsilon l}(r) \biggr],
\end{multline}
where $P_{nl}(r)$ is the radial wave function for a bound-state with quantum numbers $n$ and $l$ and $P_{\epsilon l}(r)$
is the radial wave function for a continuum state with energy $\epsilon$ and angular quantum number $l$. The
continuum states are normalized on the energy scale. 
The TF requirement Eq.(\ref{e6}) that the WS cell be neutral remains unchanged in the average-atom theory. 

\begin{figure}
\centerline{\includegraphics[scale=0.7]{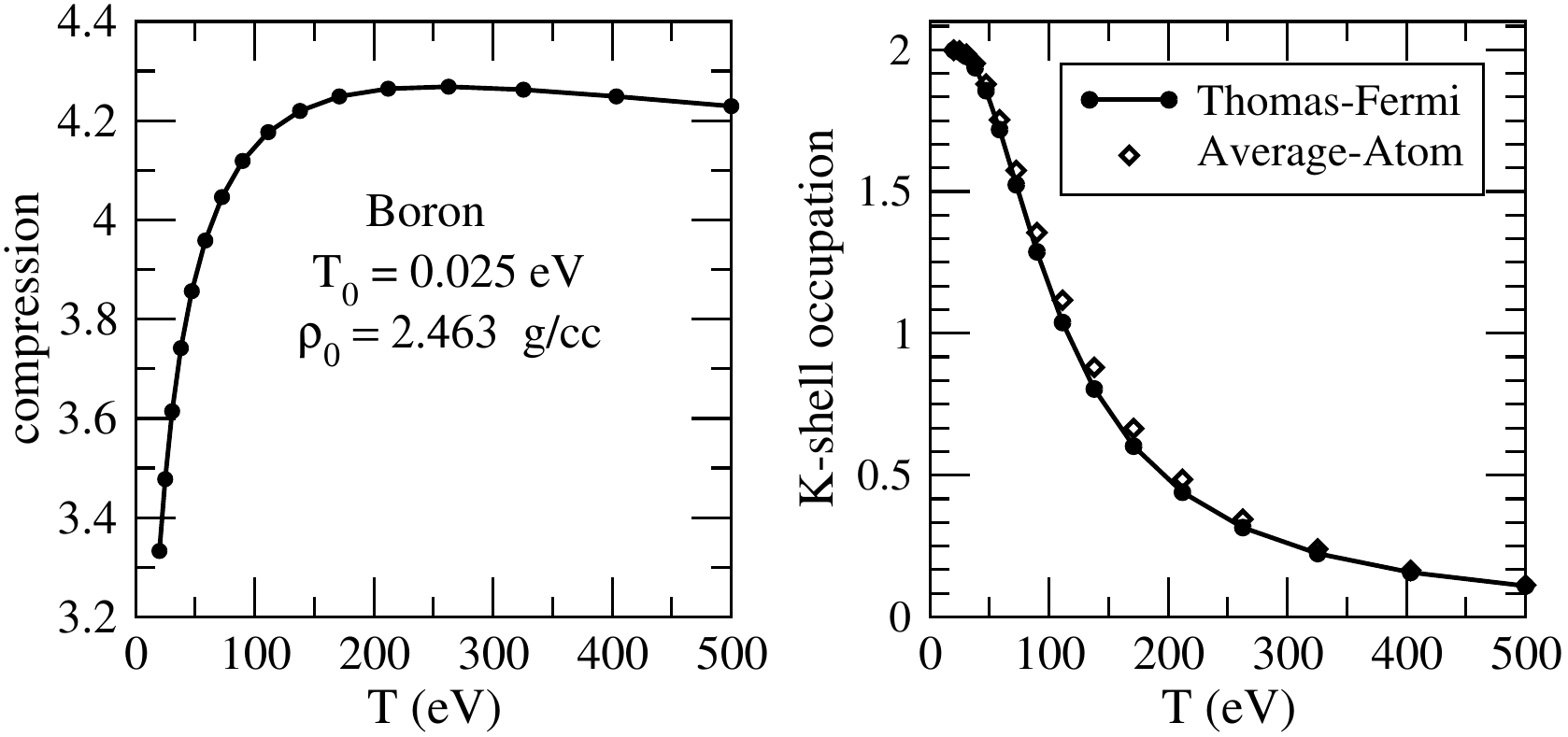}}
\caption{\label{fig4} Left panel: Hugoniot relation giving the compression ratio in the hot plasma in terms of $T$ (eV). Dots designate comparison temperatures. Right panel: K-shell occupation from the average-atom model (diamonds) compared with values from the TF theory (dots).}
\end{figure}

The K-shell occupation predicted by the average-atom model (the number of bound electrons for the cases considered here
where only the K-shell is occupied) is compared with values from the TF theory in Fig.~\ref{fig4} for temperatures ranging from 20~eV to 500~eV in the shock-heated plasma. In the left panel of Fig.~\ref{fig4}, the Hugoniot curve relating the compression ratio to the temperature behind the shock is shown. The closed circles designate specific comparison temperatures. In the right panel, values of the K-shell occupation obtained from the TF theory are shown by the black curve and the dots. These TF values are compared with occupation numbers from the average-atom model which are  marked by diamonds. The maximum relative difference between the two determinations is 10\% and occurs near 170~eV.  Other plasma properties, such as electron density and chemical potential, calculated in the average-atom approximation agree within 10\% with the corresponding
TF values.    It is found, however, that calculating opacities using  bound and continuum wave functions evaluated in the TF potential  leads to unphysical oscillations at high photon energies;  therefore, the average-atom approximation is used in the following sections to study opacities.

\section{Opacity: Atomic Scattering Factor}
An electromagnetic plane wave traveling in the $z$ direction that scatters from an ion acquires an outgoing spherical component 
\begin{equation}
 e^{i k z - i \omega t}  \to  e^{i k z - i \omega t} + \frac{r_0}{r} e^{ikr-i\omega t} f(\theta) \cos\phi ,
\end{equation}
where $\theta$ is the scattering angle and $f(\theta)$ defines the atomic scattering factor (a.k.a. scattering amplitude). 
In the above, $r_0$ is the classical electron radius, introduced to make the scattering factor dimensionless. 
The coefficient $\cos{\phi}$ ranges from 1, when the 
polarization directions of the incident and scattered wave are parallel, to 0 when they are perpendicular. 

From Huygens's principle, it follows that the scattered components of a plane wave traveling in a medium 
add coherently to create a plane wave moving in the same direction but with index of refraction $n$:
\begin{equation} 
 n = 1 - \frac{r_0 \lambda^2 }{2\pi} n_i f(0),
\end{equation}
where $\lambda$ is the wavelength of the electromagnetic wave, $n_i$ is the ion density and $f(0)$ is the forward scattering factor. The forward scattering factor $f(0)$, which is complex, is written
\begin{equation}
f(0) = f_1 -i f_2.
\end{equation}
As shown in Ref.~\cite{hen:93}, for x-ray energies greater than the photoionization threshold,
\begin{align}
f_1 &\ \approx Z, \\
f_2 & \ = \frac{\sigma_\text{bf}(\omega)}{2r_0 \lambda}, \label{f2}
\end{align}
where $\sigma_\text{bf}(\omega)$ is the bound-free photoionization cross section. 
The relation between $f_2$, which is responsible for plasma opacity, and $\sigma_\text{bf}(\omega)$ is a consequence of the optical theorem. 
It follows that the intensity of a plane electromagnetic wave traveling in the medium described above is
\begin{equation}
I(z) = I_0 \exp(-2r_0\lambda n_i f_2\, z) = I_0 \exp(-n_i\, \sigma_\text{bf}\, z).
\end{equation}

The quantum mechanical expression for photoionization of the K-shell of an atom or ion is
\begin{equation}
\sigma_\text{bf}(\omega) = \frac{8 \pi^2}{3} \alpha \omega |D|^2, \label{sig}
\end{equation}
where $D$ is the dipole matrix element
\begin{equation}
D = \int_{0}^{\infty}\!\! P_{\epsilon 1}(r) r P_{1 s}(r) dr.
\end{equation}
In the present study, we evaluate the cross section using average-atom wave functions.
To account for the fact that the K-shell is partially occupied, the cross section must be multiplied by the fractional
K-shell occupation: occ/2.
\begin{figure}
\centerline{\includegraphics[scale=0.7]{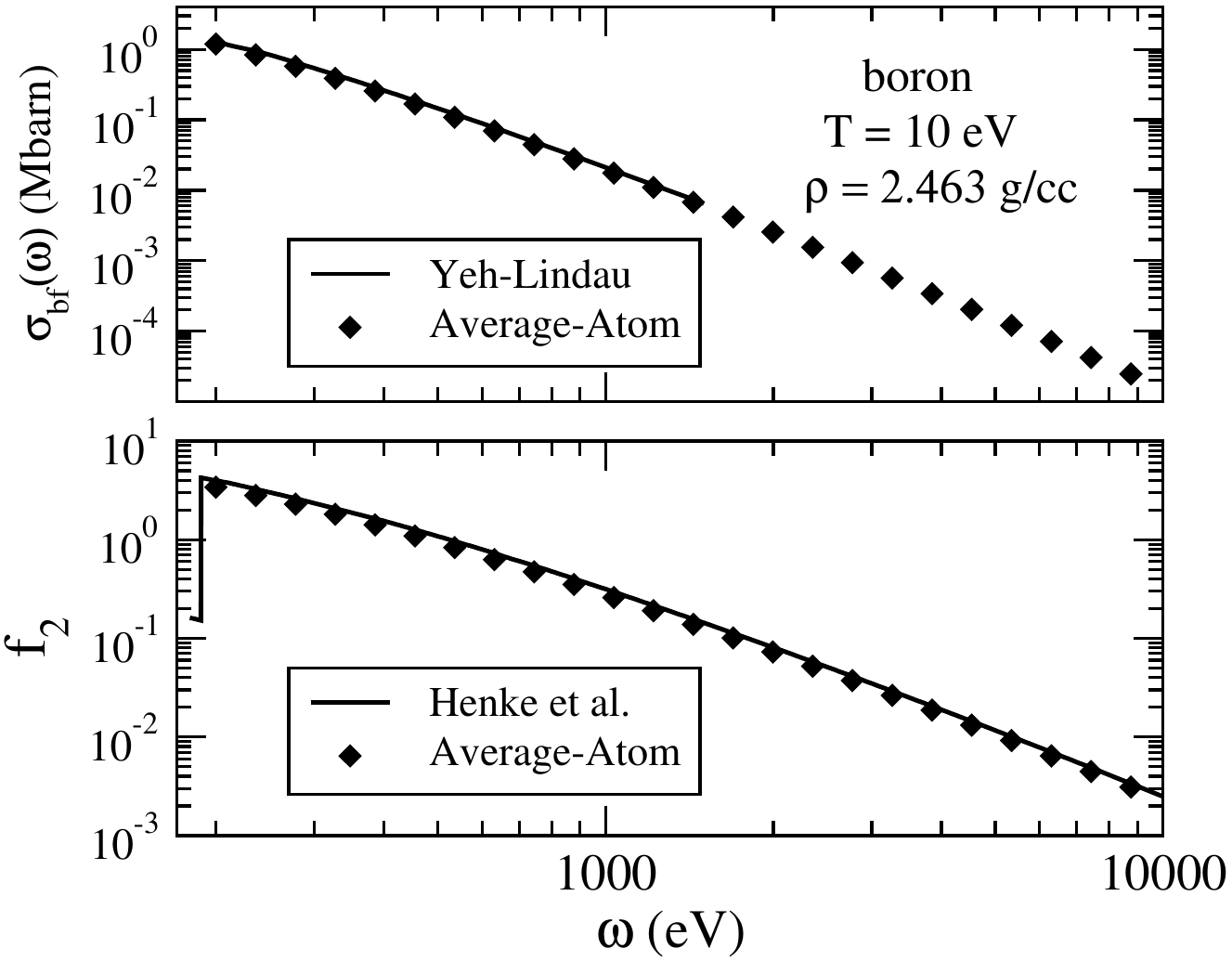}}
\caption{\label{fig5} Upper panel: The average-atom K-shell photoionization cross section for a warm-dense boron plasma at $T=10$~eV and $\rho=2.463$~g/cc in which the K-shell is fully occupied, shown by diamond symbols, is compared with the K-shell cross section for a neutral boron atom \cite{yeh:85} shown in the black line. Lower panel: The average-atom atomic scattering factor $f_2$ for warm-dense boron, shown by the diamond symbols, is compared with $f_2$ for cold boron from the Henke table \cite{hen:93}, shown in the black line.}
\end{figure}

It is of interest to compare the photoionization cross section calculated using average atom wave functions with the
corresponding neutral-atom cross section. To this end, the average-atom cross section for boron at $T=10$~eV and density $\rho=2.463$~g/cc, where the K-shell is fully occupied, is compared in the upper panel of Fig.~\ref{fig5} with the K-shell cross section for a neutral boron atom given by Yeh and Lindau \cite{yeh:85}. In the lower panel of Fig.~\ref{fig5}, the value of $f_2$ obtained from Eq.~(\ref{f2}) is compared with the cold-matter value tabulated in Ref.~\cite{hen:93}. Both the average-atom cross section and scattering factor for this warm-dense plasma, where the K-shell is fully occupied, are found to be in close agreement with well-established cold matter results.    

\begin{figure}
\centerline{\includegraphics[scale=0.65]{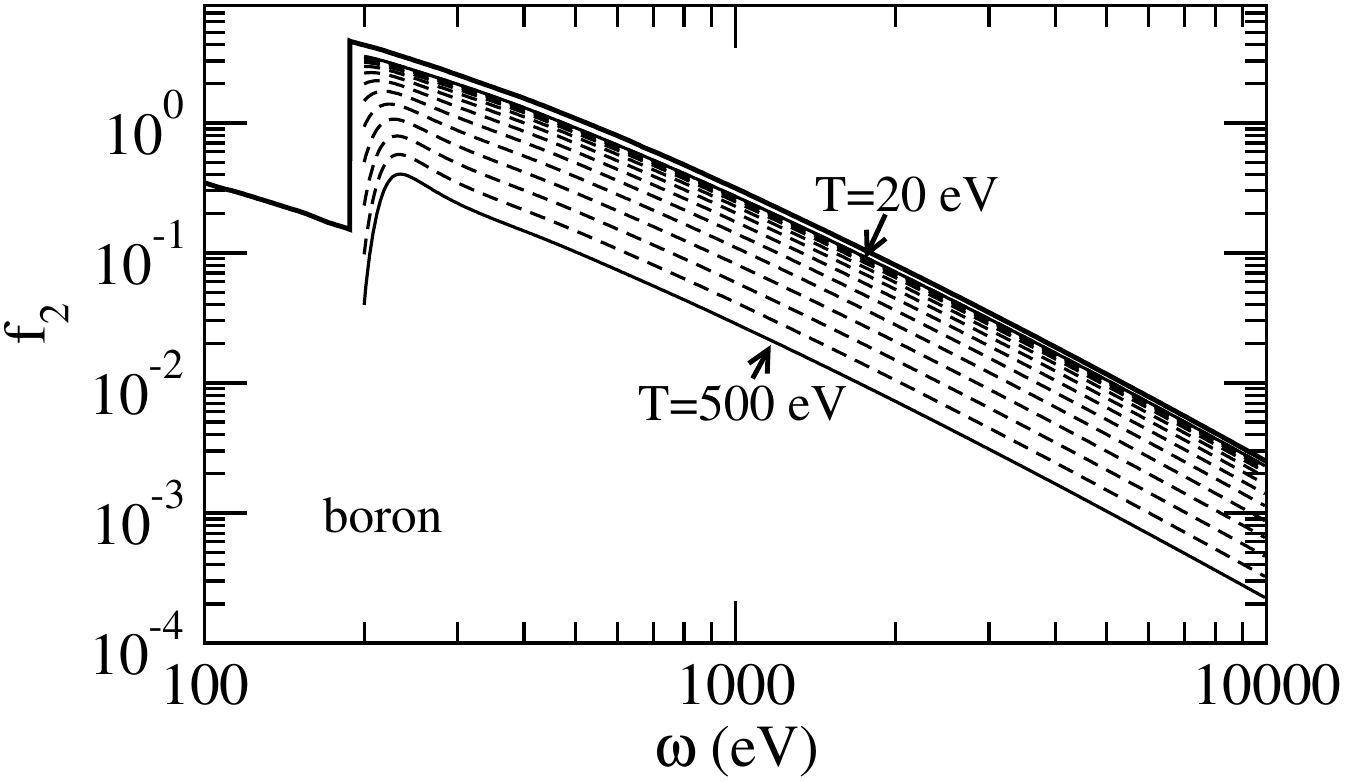}}
\caption{\label{fig6} The (cold-matter) atomic scattering factor $f_2$ for boron from Ref.~\cite{hen:93} (thick black curve) is compared with average-atom calculations of $f_2$ (dashed curves) carried out at temperatures between 20 and 500 eV  marked by the dots on the boron Hugoniot curve shown in the left panel of Fig.~\ref{fig4}}.  
\end{figure}

This situation changes as the temperature increases on the hot side of the plasma and the K-shell occupation decreases.
In Fig.~\ref{fig6}, we compare the Henke cold-matter result for $f_2$, shown by the black curve, with average-atom results for $f_2$, shown by the dashed curves. These curves are evaluated at the densities and temperatures marked out by the dots on the Hugonoit curve shown in the left panel of Fig.~\ref{fig4}. As expected, the average-atom values of $f_2$ decrease systematically with temperature along the Hugoniot.    

\section{Opacity: Mass Attenuation Coefficient}

An alternative measure of opacity is the x-ray mass attenuation coefficient $\mu/\rho$ defined and tabulated in Ref.~\cite{hubb:95}.
For single-ion plasmas, in the photon energy interval $10^2\ \text{eV}\ \leq \omega \leq 10^4\ \text{eV}$, where the x-ray opacity is governed by K-shell photoionization, the attenuation coefficient is defined by $\mu = n_i \sigma_\text{bf}$ for a single-ion plasma, leading
to the result,
\begin{equation}
I(z) = I_0 \exp(-\mu z),
\end{equation}
for the reduction of intensity of an electromagnetic wave.  
It follows from the fact that the ion number density is related to the plasma density by
\begin{equation}
n_i = \frac{N_A}{A} \rho 
\end{equation}
that
\begin{equation}
 \mu/\rho = \frac{N_a}{A} \sigma_\text{bf}
\end{equation}
depends on plasma properties only indirectly through the photoionization cross section.

For multi-ion plasmas, the expression for $\mu/\rho$ is
\begin{equation}
 \mu/\rho =  \sum_i x_i \left(\mu/\rho \right)_i,
\end{equation}
where $x_i$ is the concentration of ionic species $i$.

\begin{figure}
\centerline{\includegraphics[scale=0.7]{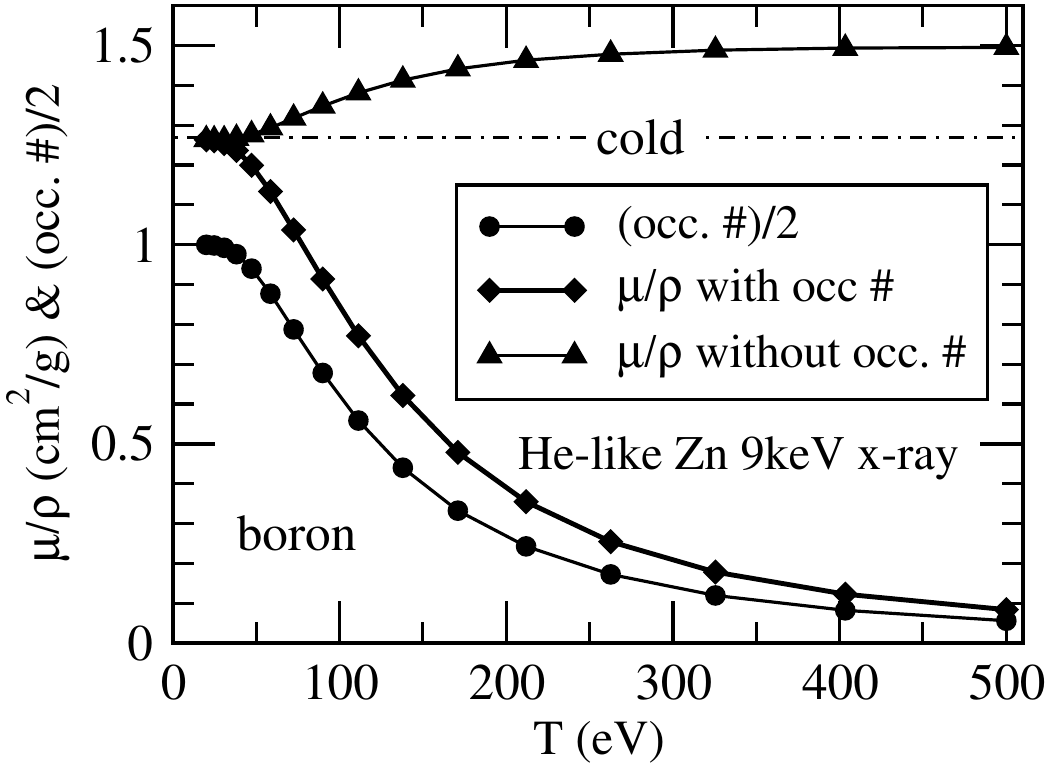}}
\caption{\label{fig7} The boron  mass attenuation coefficient $\mu/\rho$ for a 9~keV He-like Zn x-ray and the relative K-shell occupation are plotted as functions of temperature along the boron Hugoniot. The  circles give the relative K-shell occupation at points marked on the boron Hugoniot curve in Fig.~\ref{fig4}; the diamonds show $\mu/\rho$ including the K-shell occupation, whereas, the  triangles show the $\mu/\rho$ evaluated assuming a filled-K-shell. The occupation factor accounts for a 95\% reduction in the opacity over the 20-500~eV interval, while the photoionization cross section is responsible for an 18\% increase.}
\end{figure}

 The relative importance of the K-shell occupation on the $\mu/\rho$ for a 9~keV x-ray is illustrated
 in Fig.~\ref{fig7}. The curve marked by diamonds shows results of a calculation of $\mu/\rho$  in which the photoionization cross section given in Eq.~(\ref{sig}) is
multiplied by occ./2, shown by dots, accounting for a 95\% reduction of the opacity over the 20-500~eV temperature interval. The curve in Fig.~\ref{fig7} marked by triangles shows that $\mu/\rho$ without the K-shell occupation factor actually increases by 18\%. This example illustrates the importance of understanding of the K-shell occupation, but adds the caution that the photoionization cross section also varies along the Hugoniot.

\begin{figure}
 \centerline{\includegraphics[scale=0.58]{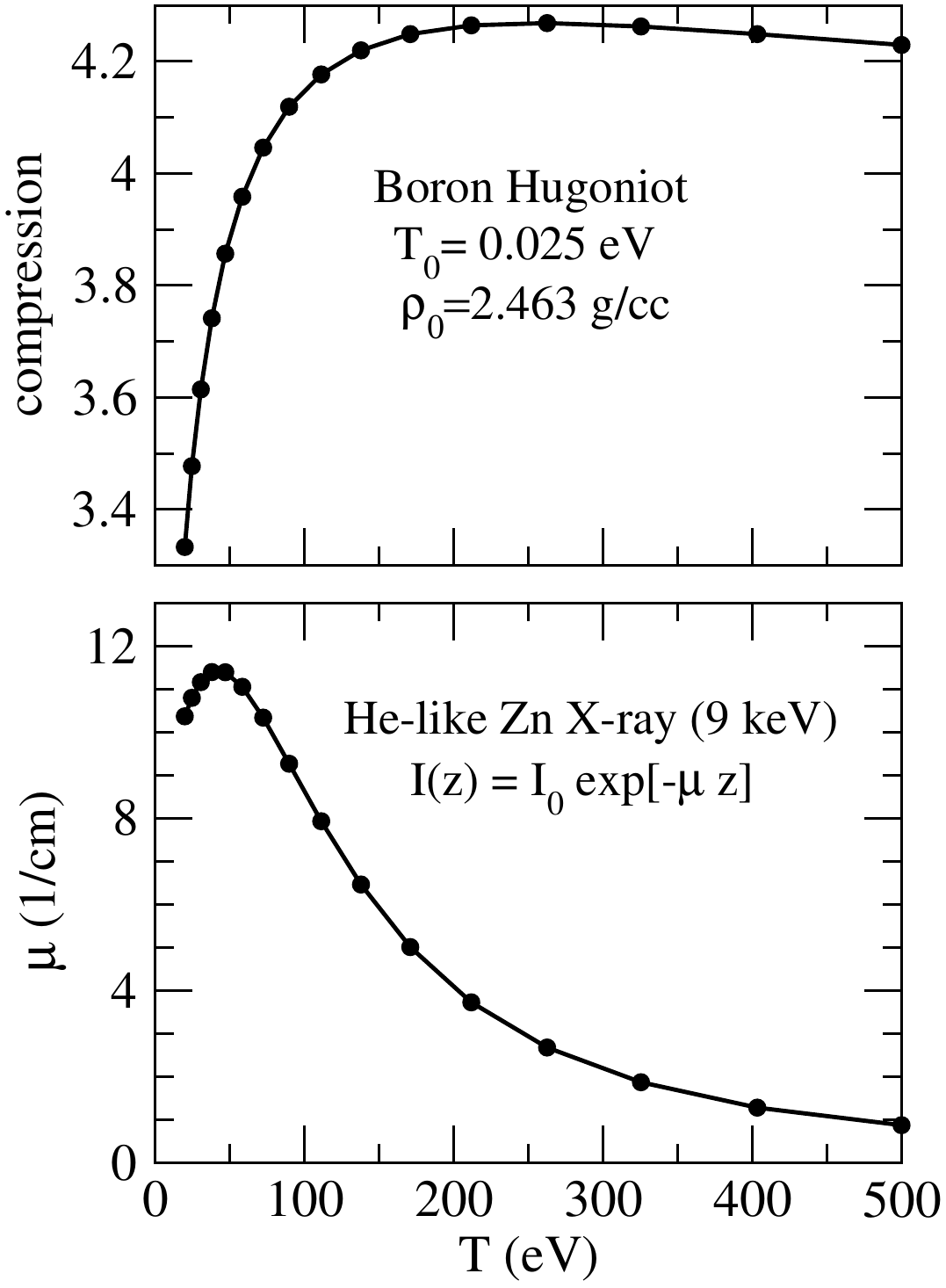}\qquad
  \includegraphics[scale=0.58]{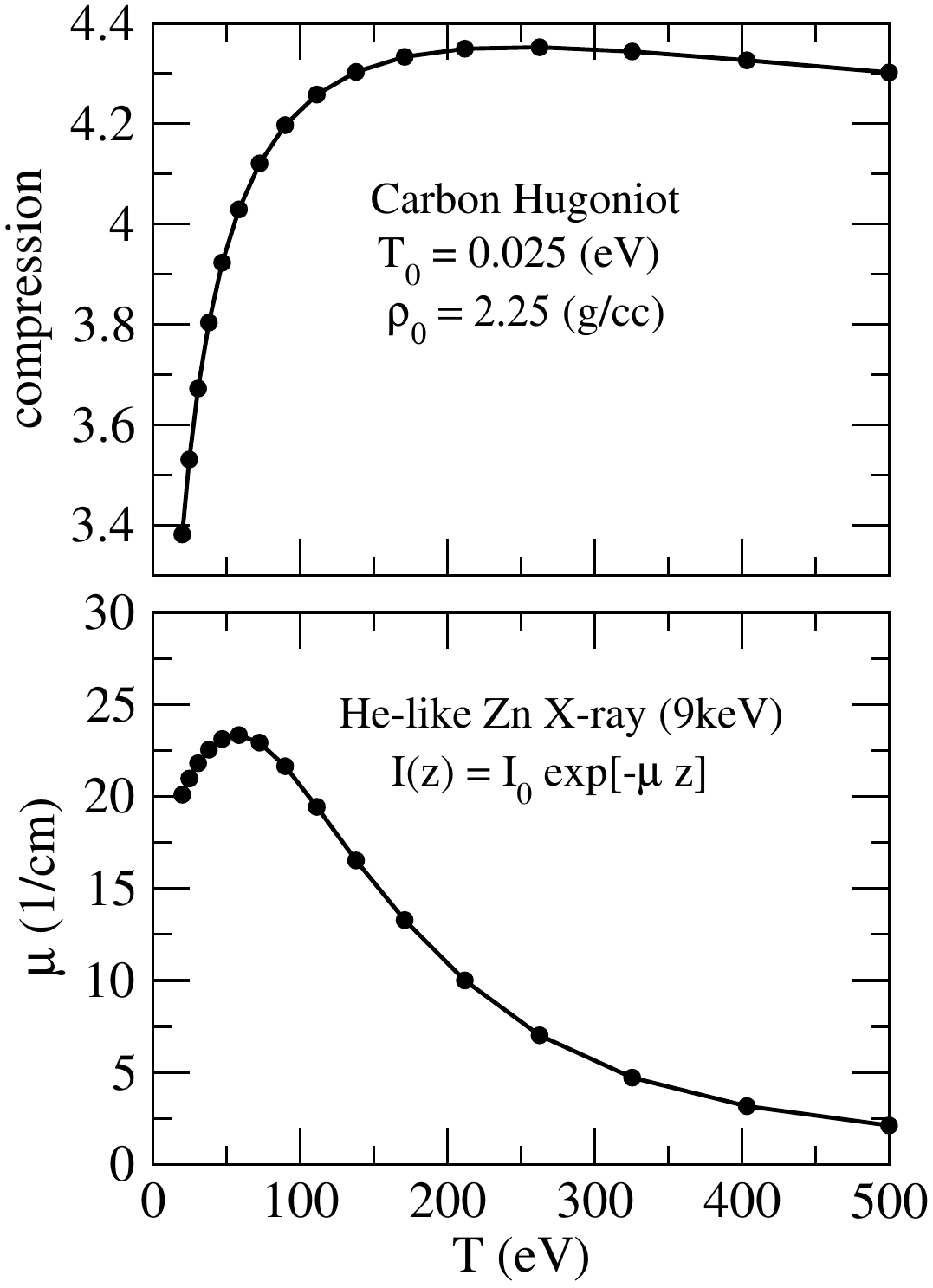}}
\vspace{2ex}

\centerline{\includegraphics[scale=0.58]{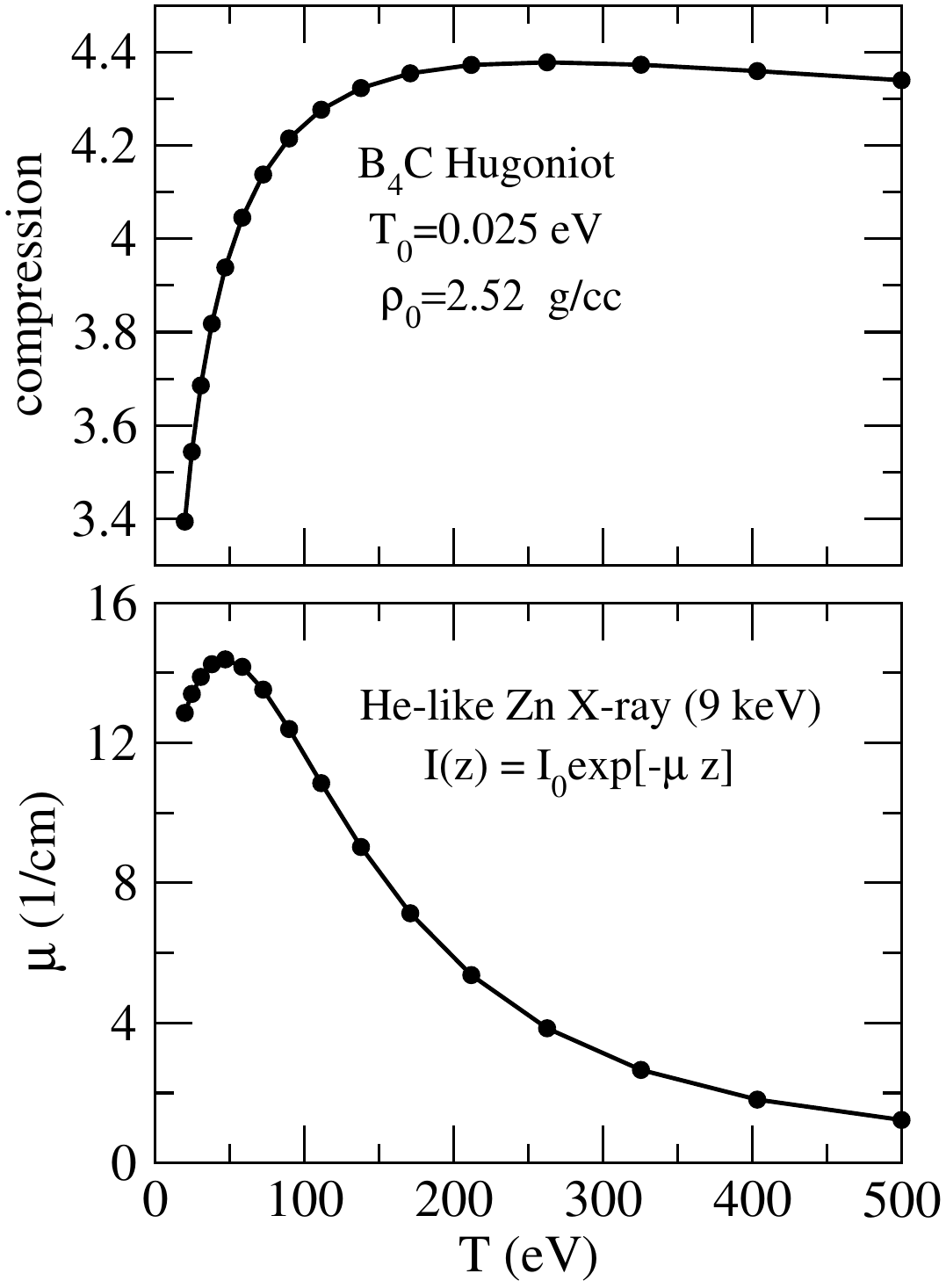} \qquad
\includegraphics[scale=0.58]{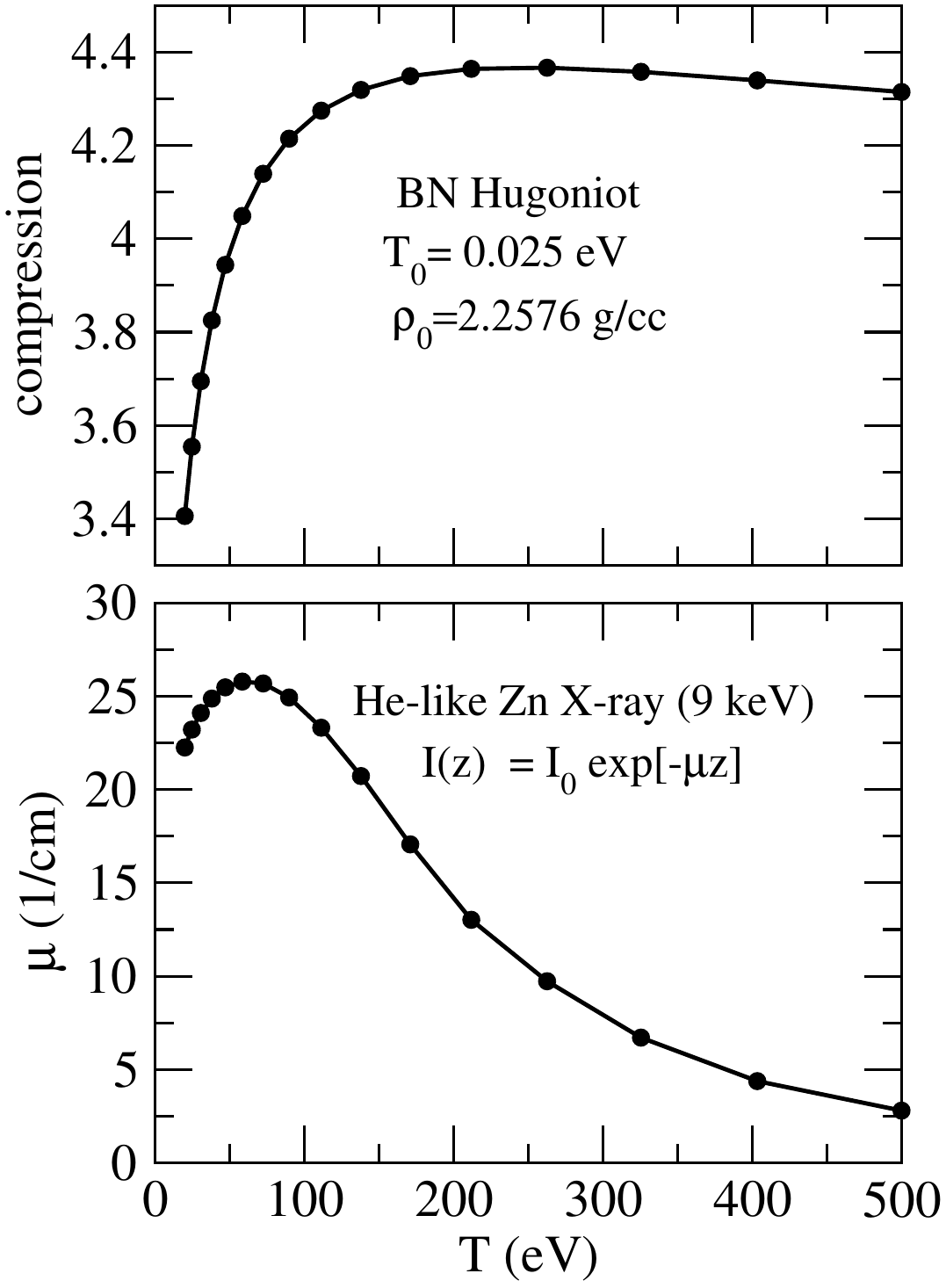}}

\caption{\label{fig8} Segments of the temperature Hugonot are shown along with the x-ray mass attenuation coefficient $\mu$  for B, C, B$_4$C and BN at points along the temperature Hugoniot. Material densities $\rho$ can be inferred from the compression ratios. The relative magnitude of $\mu(T)$ for these materials is a reflection of the strong $Z$ dependence  of the K-shell photoionization cross section.}
\end{figure}

In Fig.~\ref{fig8}, segments of the temperature Hugoniot curves between 20 and 500~eV and the corresponding x-ray mass attenuation coefficient $\mu(T)$ for a 9~keV x-ray impinging on a shock-heated plasma at temperature $T$ and compression ratio $\rho/\rho_0$ illustrate the theory developed herein for materials B, C, B$_4$C and BN mentioned in Ref.~\cite{ZM:18}. 

Densities $\rho_1$ and $\rho_2 $ of component ions in a two-component plasma with concentrations $x_1$ and $x_2$ and
density $\rho$ are determined by requiring that the WS volume of an average ion in the plasma is the average of the WS volumes of the constituent ions:
\begin{equation} 
\frac{ x_1 A_1+ x_2 A_2}{N_A \rho}= x_1\frac{ A_1}{N_A \rho_1} +  x_2\frac{A_2}{N_A \rho_2}, 
 \end{equation}
where $A_1$ and $A_2$ are the atomic weights of the ions and $N_A$ is Avogadro's constant.
Additionally, we require that the free-electron densities associated with distinct ions  be identical: 
\begin{equation}
  n_e = n_e(1) = n_e(2).
\end{equation}
Solving the two equations above simultaneously permits one to determine $\rho_1$ and $\rho_2$.  
It should be noted that the second condition also ensures that the pressure $p$ and chemical potential $\mu$ associated with each ion type, and consequently of the plasma, is unique. The electron energy in the plasma is just the weighted sum of the electron energies of the two components. 

The function $\mu(T)$ shown in Fig.~\ref{fig8} increases with temperature near $T=20$~eV because $\mu/\rho$ takes on its constant cold-matter value while $\rho$ increases with $T$ along the Hugioniot for low temperatures. 
The peak value of $\mu(T)$ for B$_4$C is slightly larger than that for B owing to the 20\% admixture of C.  Moreover, the peak values of $\mu(T)$ for BN is about twice as large as that for B and slightly larger than the peak value for C owing to the 50\% admixture of N.  The observed sensitivity of the peak values of $\mu(T)$ to ionic constituents can be traced to the strong ($Z^5$) dependence of the photoionization cross section on ionic charge $Z$.

\section{Summary and Conclusions} A systematic method for evaluating the opacity of shock-heated light-element plasmas  has been developed and applied to boron, carbon, boron carbide and boron nitride. This method consists of evaluating the shock Hugoniot in the temperature interval where the K-shell occupation drops from two to near zero and evaluating the photoionization cross section for temperatures and densities along the Hugoniot in the average-atom approximation. For (relatively) cold plasmas, where the K-shell is fully occupied, the average-atom theory leads to values of $f_2$ and $\mu/\rho$ very close to those tabulated in Refs.~\cite{hen:93,hubb:95}. As temperature increases along the Hugoniot, the opacity decreases from the cold matter values. This decrease is found to be governed dominantly by the falloff of the K-shell occupation, but to a lesser extent by the dependence of the $\sigma_\text{bf}$ on density and temperature along the Hugoniot.

\section*{Acknowledgments}
The authors owe a debt of gratitude to Phil Sterne for an introduction to Hugoniot calculations in plasmas, to Brian Wilson for assistance on average-atom calculations and to
Heather Whitley for advice on the equations of state for boron and boron compounds.
The work of J.N.\ was performed under the auspices of the U.S.\ Department of Energy by Lawrence Livermore National Laboratory under Contract DE-AC52-07NA27344.


\end{document}